\documentclass[aps,prl,twocolumn,floatfix,showpacs,superscriptaddress]{revtex4}

\usepackage{amssymb}

\usepackage{graphicx}
\usepackage{dcolumn}
\usepackage{bm}
\usepackage{color}
\begin{document}

\title{First-principles investigation of the very large Perpendicular Magnetic Anisotropy at Fe$|$MgO Interfaces}
\date{\today}
\author{H. X.~Yang}
\affiliation{SPINTEC, UMR 8191 CEA-INAC/CNRS/UJF-Grenoble 1/Grenoble-INP, Grenoble, 38054, France}
\author{J.~H.~Lee}
\affiliation{SPINTEC, UMR 8191 CEA-INAC/CNRS/UJF-Grenoble 1/Grenoble-INP, Grenoble, 38054, France}
\affiliation{Korea Institute of Science and Technology, Seoul 136-791, Korea}
\author{M.~Chshiev}
\affiliation{SPINTEC, UMR 8191 CEA-INAC/CNRS/UJF-Grenoble 1/Grenoble-INP, Grenoble, 38054, France}
\author{A.~Manchon}
\affiliation{SPINTEC, UMR 8191 CEA-INAC/CNRS/UJF-Grenoble 1/Grenoble-INP, Grenoble, 38054, France}
\affiliation{King Abdullah University of Science and Technology, Thuwal 23955-6900, Saudi Arabia}
\author{K.~H.~Shin}
\affiliation{Korea Institute of Science and Technology, Seoul 136-791, Korea}
\author{B.~Dieny}
\affiliation{SPINTEC, UMR 8191 CEA-INAC/CNRS/UJF-Grenoble 1/Grenoble-INP, Grenoble, 38054, France}

\begin{abstract}
The perpendicular magnetic anisotropy (PMA) arising at the interface between ferromagnetic transition metals and metallic oxides are investigated via first-principles calculations. In this work very large values of PMA up to 3 erg/cm$^2$ at Fe$|$MgO interfaces are reported in agreement with recent experiments. The origin of PMA is attributed to overlap between O-$p_z$ and transition metal $d_{z^2}$ orbitals hybridized with $d_{xz(yz)}$ orbitals with stronger spin-orbit coupling induced splitting around the Fermi level for perpendicular magnetization orientation. Furthemore, it is shown that the PMA value weakens in case of over- or underoxidation when oxygen $p_z$ and transition metal $d_{z^2}$ orbitals overlap is strongly affected by disorder, in agreement with experimental observations in magnetic tunnel junctions.
\end{abstract}
\maketitle

Spin-orbit interaction (SOI) plays a major role in a wide class of physical phenomena both from fundamental and applications points of view \cite{ReviewSpin}. For instance, it is at the heart of basic magnetic phenomena such as magnetocrystalline anisotropy \cite{TAMRPRB}, Rashba effect \cite{Rashba, reviewRashba} and magnetization damping. Controlling SOI strength at the interface between ferromagnetic (FM) and non-magnetic layers represents an outstanding challenge for advancement of transport and magnetic properties of spintronic magnetic devices, such as perpendicular Magnetic Tunnel Junctions~\cite{pMTJ1,pMTJ2,pMTJ3,pMTJ4,pMTJ5,Lavinia} (p-MTJs) and tunneling anisotropic magnetoresistive (TAMR) systems~\cite{TAMR1}. Recently, electric field control of interfacial magnetic anisotropy has attracted much attention as well~\cite{Suzuki,Nakamura}. Traditionally, interfaces between magnetic and heavy non-magnetic transition metals such as Co$|$Pt~\cite{nakajima}, Co$|$Pd~\cite{Pd1,Pd2}, Co$|$Au~\cite{weller} have been used to obtain perpendicular magnetic anisotropy (PMA). It has been shown that the onset of the PMA at these interfaces is related to an increase of the orbital momentum of Co~\cite{weller} due to the strong hybridization between the 3$d$ orbitals of the transition metal and the 5$d$ orbitals of heavy metal~\cite{nakajima}. This hybridization enhances the energy splitting between the Co 3$d_{z^2}$ and 3$d_{x^2-y^2}$ orbitals and induces a charge transfer between the two layers~\cite{bruno, daalderop,daalderop1}. As a result, the combination between SOI and hybridization-induced charge transfer leads to the PMA. Thus, the presence of a heavy non-magnetic layer (Pt, Pd, Au, W, Mo) was believed to be essential to trigger the PMA. 
\begin{figure}
  \includegraphics[width=6.618 cm]{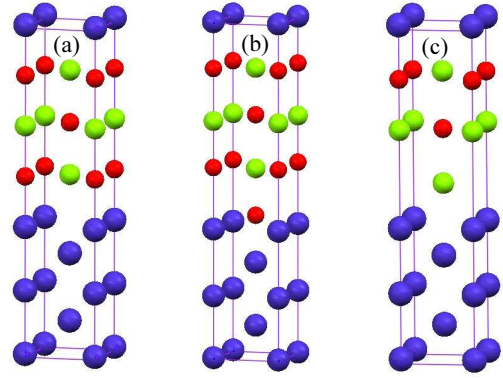}\\
   \caption{Schematics of the calculated crystalline structures for (a) pure, (b) over-oxydized and (c) under-oxydized geometries. Fe, Mg and O are represented by blue, green and red balls respectively.}\label{fig1}
\end{figure}

However, Monso et al have shown that PMA could be observed also at Co(Fe)$|$MOx interfaces (M=Ta, Mg, Al, Ru etc.)~\cite{monso} in spite of the weak SOI at the interface. Surprisingly large PMA values up to 1 to 2 erg/cm$^2$ have been reported, which are comparable or even larger than the PMA observed at Co$|$Pt or Co$|$Pd interfaces~\cite{guo}. 
This result is quite general and has been observed in both crystalline (MgO) or amorphous (AlOx) barriers, using both natural or plasma oxidation~\cite{jap}. The PMA could be dramatically improved under annealing~\cite{rodmacq,Lavinia} and X-ray Photoemission Spectroscopy has demonstrated that the PMA could be correlated without ambiguity with the presence of oxygen atoms at the interface~\cite{jap}. In fact, a correlation between PMA and oxidation conditions have been demonstrated for a wide range of FM$|$MOx including those based on Co$_{x}$Fe$_{1-x}$, thus indicating that the phenomenon is quite general at interfaces between magnetic transition metals and oxygen terminated oxides. These observations led the authors to postulate that, in spite of the weak SOI of the elements (Fe, Co, Al, O), oxidation condition plays an essential role in the PMA, as it does in TMR~\cite{tmrOxidation} or interlayer exchange coupling (IEC)~\cite{MgOIEC}. Recent experiments reported large PMA values of 1.3~erg/cm$^2$ at CoFeB$|$MgO structures~\cite{Ohno,MgOPMAAPLohno}. Furthermore, it has been experimentally demonstrated that there is a strong correlation between PMA and TMR maximum values obtained at the same optimal oxidation and annealing conditions~\cite{Lavinia2}.   
\begin{figure}
  \includegraphics[width=8.5cm]{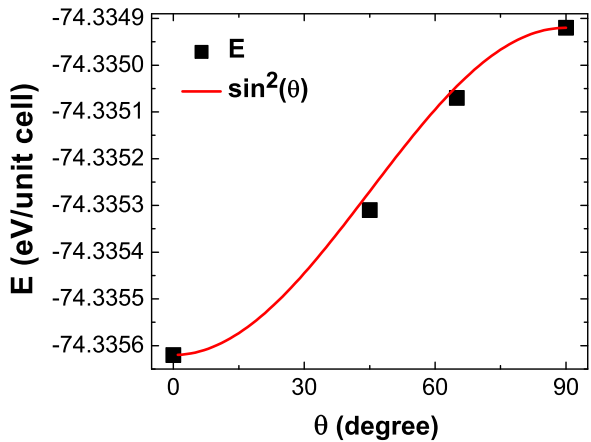}\\
   \caption{Angular dependence of the magnetic energy, where $\theta$ is the angle between the magnetization direction and the normal to the interface plane.}\label{fig2}
\end{figure}

In this letter, we report first-principles investigations of the PMA and the effect of interfacial oxidation conditions on the PMA at Fe$|$MgO(100) structures. The latter can be viewed as a model system for FM$|$MOx interfaces involving bcc electrodes including Co$_{x}$Fe$_{1-x}$ alloys.  In agreement with experiments, it is demonstrated despite the weak SOI, the bonding between the Fe-3$d$ and O-2$p$ orbitals gives rise to even stronger PMA compared to that of Co$|$Pt interfaces. The largest PMA value is obtained for ideal interfaces while it is reduced for the case of over- or underoxidized interfaces.

For the $ab~inito$ calculations, Vienna $ab ~initio$ simulation package (VASP)~\cite{vasp} was used with  generalized gradient approximation~\cite{gga} and projector augmented wave potentials~\cite{paw}.  The calculations were performed in three steps. First, full structural relaxation in shape and volume was performed until the forces are smaller than 0.001 eV/\AA ~for determining the most stable interfacial geometries. Next, the Kohn-Sham equations were solved with no spin-orbit interaction taken into account to find out the charge distribution of the system's ground state. Finally, the spin-orbit coupling was included and the total energy of the system was determined as a function of the orientation of the magnetic moments. A 19$\times$19$\times$3 K-point mesh was used in our calculations with the energy cut-off equal to 520~eV.
Three structures were considered as shown in Fig.~\ref{fig1}: (a) "pure" (O-terminated) interface, (b) over-oxidized interface (with O inserted at the interfacial magnetic layer), and (c) under-oxidized (Mg-terminated) interface. We point out that the situation of "pure" interface is the most stable one as observed in annealing experiments~\cite{rodmacq}. The most stable location for the oxygen atoms is on top of metal ions due to strong overlap between Fe-3$d$ and O-2$p$ orbitals. Correlatively, it is intersting to note that this structural configuration also yields the spin filtering phenomenon based on Bloch states symmetry leading to high TMR values~\cite{ButlerMgOTMR,MathonUmerski}. Furthermore, the strong hybridization significantly modifies the band structure giving rise to a high interfacial crystal field~\cite{bruno}.

\begin{figure}
  \includegraphics[width=8 cm, height=6.3 cm]{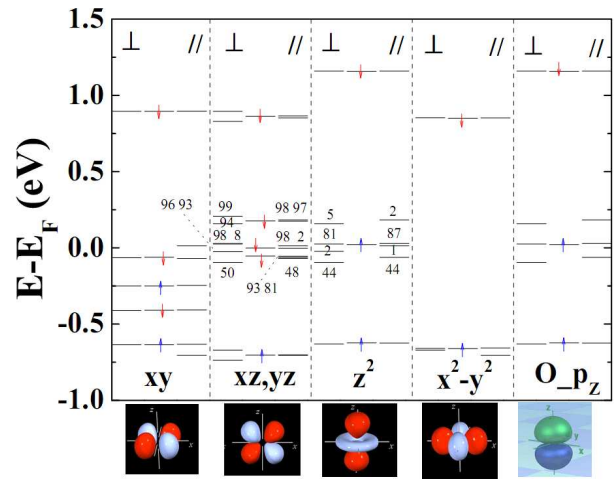}\\
   \caption{Spin-orbit coupling effects on wave function character at $\Gamma $ point of interfacial Fe $d$ and neighbor oxygen $p_z$ orbitals for pure Fe$|$MgO interface shown in Fig.~\ref{fig1}a. Three subcolumns in each column show the band levels for out-of-plane(left) and in-plane(right) orientation of magnetization as well as for the case with no spin-orbit interaction included(middle), respectively. Numbers are the percentage of the orbital character components within Wigner-Seitz spheres around interfacial atoms.}\label{fig3}
\end{figure}

In Fig.~\ref{fig2}, we present the calculated energy per unit cell as a function of the angle $\theta$ between magnetization orientation and normal to the plane for pure unrelaxed Fe$|$MgO interface. The dependence is well fitted by the conventional uniaxial anisotropy formula $E_{A}=K_0+K_2\sin^2\theta$, where $K_2=0.7$~meV/atom ($K_2=1.36$~erg/cm$^2$). Interestingly, the perpendicular surface anisotropy in this case is stronger than that of Co$|$Pt~\cite{guo} in agreement with recent experiments~\cite{Ohno}. The calculated anisotropy value is further enhanced for relaxed structures as shown in Table~\ref{Table1}. The PMA for relaxed structures weakens in case of interfacial disorder and becomes equal to 2.27 and 0.93~erg/cm$^2$ for under- and overoxidized cases, respectively (see Table~\ref{Table1}), indicating that the oxidation condition plays a critical role in PMA as it does in both TMR~\cite{tmrOxidation} and IEC~\cite{MgOIEC}. Furthermore, the tendency of PMA to decrease with oxygen excess or deficit along the metal/oxide interface is consistent with the recent experimental observations of PMA dependence on annealing temperature and oxydation conditions~\cite{Lavinia,Lavinia2}. It was reported that with higher annealing temperatures PMA increases due to interfacial quality improvement~\cite{Lavinia}. As we stated above, PMA reaches a maximum value corresponding to the TMR ratio maximum indicating that ideal interfaces are crucial also for PMA observation~\cite{Lavinia2}.

\begin{table}[b]
\caption {PMA(in unit of erg/cm$^2$) and magnetic moment ($\mu_B$ per Fe atom) for different layers of Fe in Fe$|$MgO MTJs with different oxidation conditions. }
\label{Table1}
	\begin{tabular}{lllll}
		\hline\hline
            & &  &Fe$|$MgO  \\
                            &        & pure~~~~  &under- &over-\\ 
 \hline
         PMA  (relaxed)~~        &        & 2.93   & 2.27  & 0.98 \\ 
    \hline
                            & interfacial~~ & 2.73   & 2.14  & 3.33 \\
        Moment ($\mu_B$)             & sublayer   & 2.54   & 2.41  & 2.70 \\ 
                  & bulk   & 2.56   & 2.55  & 2.61 \\ 
     \hline\hline
   \end{tabular}
\end{table}

In Table~\ref{Table1} we give also the evolution of interfacial Fe magnetic moments as a function of distance from the interface. One can see that compared to the pure case, the moments are enhanced(weakened) for over(under)oxidized interfaces.

Let us now proceed with the explanation of the physical origin of the results obtained from first principles on the effect of oxidation conditions on PMA.  
To understand the PMA origin at Fe$|$MgO interfaces, we performed detailed analysis of the impact of spin-orbit interaction on electronic band structure with out-of-plane ($d_{z^2}$, $d_{xz}$, $d_{yz}$) and in-plane ($d_{x^2-y^2}$, $d_{xy}$) Fe-3$d$ and O-$p_z$ orbitals character. 
\begin{figure}
  \includegraphics[width=8 cm, height=6.3 cm]{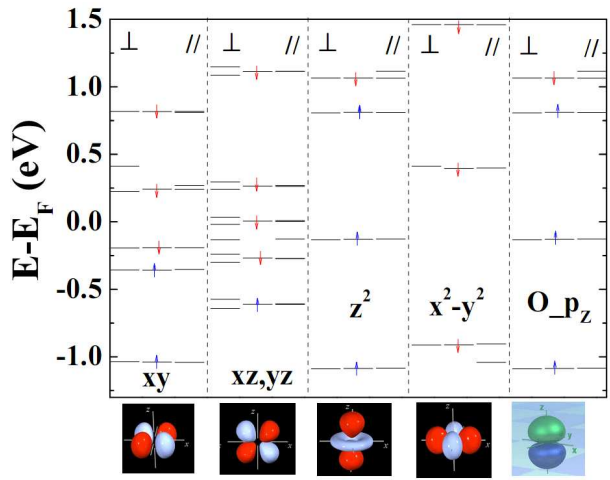}\\
   \caption{The same as Fig.~\ref{fig3} for over-oxidized Fe$|$MgO interface.}\label{fig4}
\end{figure}

We start from the analysis for pure interfaces represented in Fig.~\ref{fig1}a. In Fig.~\ref{fig3}, we show bands around the Fermi level $E_F$ at $\Gamma$-point with orbital and interfacial atoms projected wave function character for out-of-plane (left) and in-plane (right) orientation of the magnetization as well as in the absence of spin-orbit coupling (middle). When no SOI is included (middle subcolumns), one can clearly see the band level resulting from hybridization between Fe-$d_{z^2}$ and O-$p_z$ orbitals. This is a signature of Bloch state with $\Delta_1$ symmetry around $E_F$ for majority Fe and MgO which is at the heart of the spin filtering phenomenon causing enhanced TMR values in MgO based MTJs~\cite{ButlerMgOTMR}. The double degenerated bands with $\Delta_5$ symmetry related to minority Fe are also present close to the Fermi level. When spin-orbit interaction is switched on, the degeneracy is lifted and majority $\Delta_1$ and minority $\Delta_5$ are mixed up producing bands with both symmetry characters. As a result, band levels with $d_{z^2}$, $d_{xz}$, $d_{yz}$ and $p_z$ character appear splitted around the Fermi level and this splitting is larger and the lowest band deeper for out-of-plane magnetization orientation as clearly seen in left and right subcolumns in Fig.~\ref{fig3}, respectively. Thus, the lift of degeneracy of $d_{xz}$ and $d_{yz}$ orbitals combined with mixing with Fe-$d_{z^2}$ and O-$p_z$ orbitals is at origin of perpendicular magnetic anisotropy for pure Fe$|$MgO interfaces. This result shows that the out-of-plane components of $d_{xz,yz}$ orbitals plays a crucial role for PMA similarly to Co$|$Pd interfaces~\cite{PMACoPd}.

Next, we proceed with the same analysis for under- and over-oxidized Fe$|$MgO interfaces represented in Fig.~\ref{fig1}(b) and (c), respectively. As shown in Fig.~\ref{fig4} for the case with an additional oxygen located at the Fe$|$MgO interface (Fig.~\ref{fig1}(b)), spin orbit coupling lifts again the degeneracy for states with $d_{xz,yz}$ causing stronger splitting and deeper level position for out-of-plane orientation of magnetization compared to the in-plane one. However, these states are not mixed anymore with Fe-$d_{z^2}$ and O-$p_z$ orbitals due to local charge redistribution induced by additional oxygen atoms~\cite{tmrOxidation}. Since $d_{z^2}$ and $p_z$ orbital hybridization which is mainly responsible for PMA is not splitted, the anisotropy is significantly reduced. 

\begin{figure}
  \includegraphics[width=8 cm, height=6.3 cm]{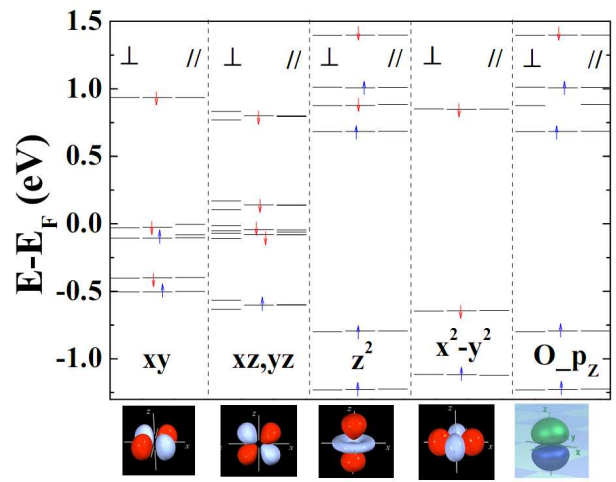}\\
   \caption{The same as Fig.~\ref{fig3} for under-oxidized Fe$|$MgO interface.}\label{fig5}
\end{figure}

\begin{figure}
  \includegraphics[width=8 cm]{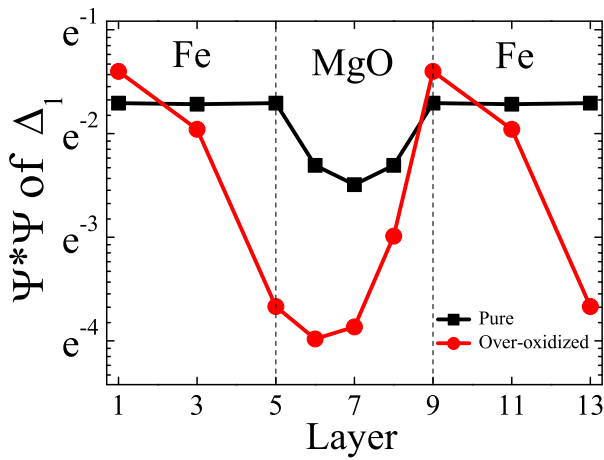}\\
   \caption{$\Delta _1$ Bloch state character at $\Gamma$-point around the Fermi level as a function of layer number in pure and over-oxidized Fe$|$MgO interfaces shown in Fig.~\ref{fig1}(a) and (b), respectively. $\Delta _1$ Bloch state is absent around Fermi level in underoxidized case shown in Fig.~\ref{fig1}(c).}\label{fig6}
\end{figure}

A different picture occurs in case of underoxidized Fe$|$MgO interface represented in Fig.~\ref{fig1}(c). As shown in Fig.~\ref{fig5}, the Fe-$d_{z^2}$ and O-$p_z$ components around the Fermi-level are now absent. As a result, the degeneracy lift induced by spin-orbit interaction for states with $d_{xz,yz}$ character is now solely responsible for the PMA. Since the splitting of these $d_{xz,yz}$ orbitals is still relatively strong and higher for out-of-plane magnetization orientation compared to the in-plane one, anisotropy values are higher compared to overoxidized case but lower compared to the ideal Fe$|$MgO interfaces. Thus, the PMA reaches its maximum for ideal interfaces. In order to understand the correlation between PMA and TMR, in Fig.~\ref{fig6} we plot the wave function character of $\Delta_1$ Bloch state as a function of the position across the supercells used for PMA calculations. One can clearly see that the $\Delta_1$ decay rate is strongly enhanced in the case of overoxidized interface compared to the ideal one. There is no $\Delta_1$ band around the Fermi level for underoxidized case as demonstrated above. This explains why both PMA and TMR reach the maximum values in a correlated way as observed experimentally ~\cite{Lavinia2}, this maximum being reached for ideal interfaces.

In conclusion, we presented $ab~initio$ studies of perpendicular magnetic anisotropy at Fe$|$MgO interfaces as a function of the oxygen content along the interface. The PMA values are highest in the case of pure interfaces in agreement with recent experimental studies~\cite{Ohno,Lavinia2} and may reach up to 3~erg/cm$^2$ for relaxed interfaces. The origin of large PMA is ascribed to the spin-orbit induced mixing between $\Delta_1$ and $\Delta_5$-like 3$d$ and 2$p$ orbitals at the interface between the transition metal and the insulator combined with the  degeneracy lift of out-of-plane 3$d$ orbitals. The PMA amplitude degrades in the case of over- or underoxidized interfaces in agreement with recent experiments~\cite{Lavinia,Lavinia2}. This is due to the impact of splitting(disappearing) of $\Delta_1$-like hybridized states around the Fermi level in presence(absence) of additional oxygen atom.

\begin{acknowledgments}
We thank L.~Nistor, B.~Rodmacq, A.~Fert, H.~Jaffres, O.~Mryasov, A.~Schuhl and W.~H.~Butler for fruitful discussions. This work was supported by Chair of Excellence Program of the Nanosciences Foundation in Grenoble, France, ERC Advanced Grant Hymagine and the KRCF DRC program. 
\end{acknowledgments}


\begin{thebibliography}{200}
\bibitem{ReviewSpin} J.~Fabian et al, Acta Phys. Slov. {\bf57}, 565 (2007).
\bibitem{TAMRPRB} A. Matos-Abiague and J. Fabian,  Phys. Rev. B {\bf79}, 155303 (2009).
\bibitem{Rashba} Yu A. Bychkov and E. I. Rashba, Journal of Physics C: Solid State Physics, {\bf17}, 6039 (1984).
\bibitem{reviewRashba}I.~Zutic et al, Review of Modern Physics {\bf76}, 323(2004).
\bibitem{pMTJ1} K. Mizunuma et al, Appl. Phys. Lett. {\bf 95}, 232516 (2009). 
\bibitem{pMTJ2} G. Kim et al, and T. Miyazaki, Appl. Phys. Lett. {\bf92}, 172502 (2008). 
\bibitem{pMTJ3}C. Ducruet et al, J. Appl. Phys. {\bf103}, 07A918 (2008). 
\bibitem{pMTJ4}J.-H. Park et al, J. Appl. Phys. {\bf103}, 07A917 (2008). 
\bibitem{pMTJ5}D. Lim et al, J. Appl. Phys. {\bf97}, 10C902 (2005). 
\bibitem{Lavinia} L. E. Nistor et al, Appl.  Phys.  Lett. {\bf 94}, 012512 (2009).
\bibitem{TAMR1} L. Gao et al, Phys. Rev. Lett. {\bf99}, 226602 (2007);  B. G. Park et al, Phys. Rev. Lett. {\bf100}, 087204 (2008). 

\bibitem{Suzuki} Y.~Shita et al., Appl. Phys. Express {\bf 2}, 063001 (2009).
\bibitem{Nakamura} K.~Nakamura et al, Phys. Rev. B {\bf 81}, 220409(R) (2010).
\bibitem{nakajima} N. Nakajima et al, Phys. Rev. Lett. {\bf81}, 5229 (1998).
\bibitem{Pd1} P. F. Carcia et al, Appl. Phys. Lett. {\bf47}, 178 (1985).
\bibitem{Pd2} H. J. G. Draaisma et al, J. Magn. Magn. Mater. {\bf66}, 351 (1987).
\bibitem{weller} D. Weller et al, Phys. Rev. B {\bf49}, 12888 (1994).
\bibitem{bruno} P. Bruno, Phys. Rev. B {\bf39}, 865 (1989).
\bibitem{daalderop} G. H. O. Daalderop et al, Phys. Rev. B {\bf50}, 9989 (1994). 
\bibitem{daalderop1}K. Kyuno et al, J. Phys. Soc. Jpn. {\bf61}, 2099 (1992).
\bibitem{monso} S. Monso et al, Appl. Phys. Lett. {\bf 80}, 4157 (2002); B. Rodmacq et al, J. Appl. Phys.{\bf 93}, 7513 (2003).

\bibitem{guo} V. W. Guo et al, J. Appl. Phys. {\bf99}, 08E918 (2006); M.~T.~Johnson et al, J. Magn. Magn. Mater. {\bf 148}, 118 (1995).

\bibitem{jap}D. Lacour et al, Appl. Phys. Lett. {\bf90}, 192506 (2007); A.~Manchon et al,  J. Appl. Phys. {\bf 104}, 043914 (2008).

\bibitem{rodmacq} A. Manchon et al, J. Magn. Magn. Mater. {\bf320}, 1889 (2008); B. Rodmacq et al, Phys. Rev. B {\bf79}, 024423 (2009)

\bibitem{tmrOxidation} X.-G. Zhang et al, Phys. Rev. B {\bf68}, 092402 (2003).
\bibitem{MgOIEC} H. X. Yang et al, Appl. Phys. Lett. {\bf96}, 262509 (2010).
\bibitem{Ohno} S. Ikeda et al, Nature Mater. {\bf9}, 271 (2010).
\bibitem{MgOPMAAPLohno}M. Endo et al, Appl. Phys. Lett. {\bf ~96}, 212503 (2010).
\bibitem{Lavinia2} L.~E. Nistor et al, Magnetics, IEEE Trans. Mag , {\bf 46}, 1412 (2010).
\bibitem{vasp}G. Kresse and J. Hafner, Phys. Rev. B {\bf47}, 558 (1993); {\bf54}, 11169 (1996); Comput. Mater. Sci. {\bf6}, 15 (1996).

\bibitem{gga}Y. Wang and J. P. Perdew, Phys. Rev. B {\bf44}, 13298 (1991).
\bibitem{paw} P.~E. Bl\"ochl, Phys. Rev. B {\bf50}, 17953 (1994); G. Kresse and D. Joubert, Phys. Rev. B {\bf59} 1758 (1999).
\bibitem{ButlerMgOTMR} W. H. Butler et al, Phys. Rev. B {\bf63}, 054416 (2001).
\bibitem{MathonUmerski} J. Mathon and A. Umerski, Phys. Rev. B {\bf63}, 220403(R) (2001).
\bibitem{PMACoPd} D.~S. Wang, R. Wu and A. H. Freeman, Phys. Rev. B {\bf48}, 15886 (1993).


\end{thebibliography}
\end{document}